\documentclass[aps,prb,showpacs,floatfix,twocolumn,superscriptaddress,longbibliography]{revtex4-2}
\usepackage{graphicx}
\usepackage{bm,amsmath,amssymb}
\usepackage{physics}
\usepackage{dcolumn}
\usepackage{subfigure}
\usepackage{color}
\usepackage{tikz}
\usepackage{fontawesome}
\usepackage{esint} 

\newcommand\link[1]{\href{#1}{\faLink}}

\usepackage{float}
\usepackage[%
    colorlinks=true,
    pdfborder={0 0 0},
    linkcolor=red,
    citecolor = red
]{hyperref}
\usepackage{ifthen}
\usepackage{mathpazo}

\makeatletter
\renewcommand{\maketag@@@}[1]{\hbox{\m@th\normalsize\normalfont#1}}%
\makeatother
\begin{document}
\title{  Unsupervised machine learning  for identifying phase transition using two-times clustering}

\author{Nan Wu}

\affiliation{College of Physics, Taiyuan University of Technology, Shanxi 030024, China}

\affiliation{
School for Physical  Sciences, University of Science and Technology of China, Hefei 230026, China}

\author{Zhuohan Li}
\affiliation{College of Software, Taiyuan University of Technology, Shanxi 030024, China}




\author{Wanzhou Zhang}
\email{Correspoinding author: zhangwanzhou@tyut.edu.cn}
\affiliation{College of Physics, Taiyuan University of Technology, Shanxi 030024, China}
\affiliation{
Hefei National Laboratory for Physical Sciences at the Microscale and Department of Modern Physics, University of Science and Technology of China, Hefei 230026, China}

\begin{abstract}
In recent years,  developing unsupervised machine learning for
identifying phase transition is a research direction. 
In this paper, we introduce a  two-times clustering method that can help select perfect configurations from a set of degenerate samples
and assign the configuration with labels in a manner of unsupervised machine learning.  
These perfect configurations can then be used to train a neural network to classify phases.
The derivatives of the predicted classification  in the phase diagram, show peaks at the phase transition points. The effectiveness of our method is tested for  the  Ising, Potts,  and Blume-Capel models.  By using the ordered configuration from two-times clustering, our method can provide a useful way to obtain  phase diagrams.
\end{abstract}

\maketitle


\begin{figure*}
\centering
\includegraphics[width=0.75\textwidth,height=0.6\textwidth]{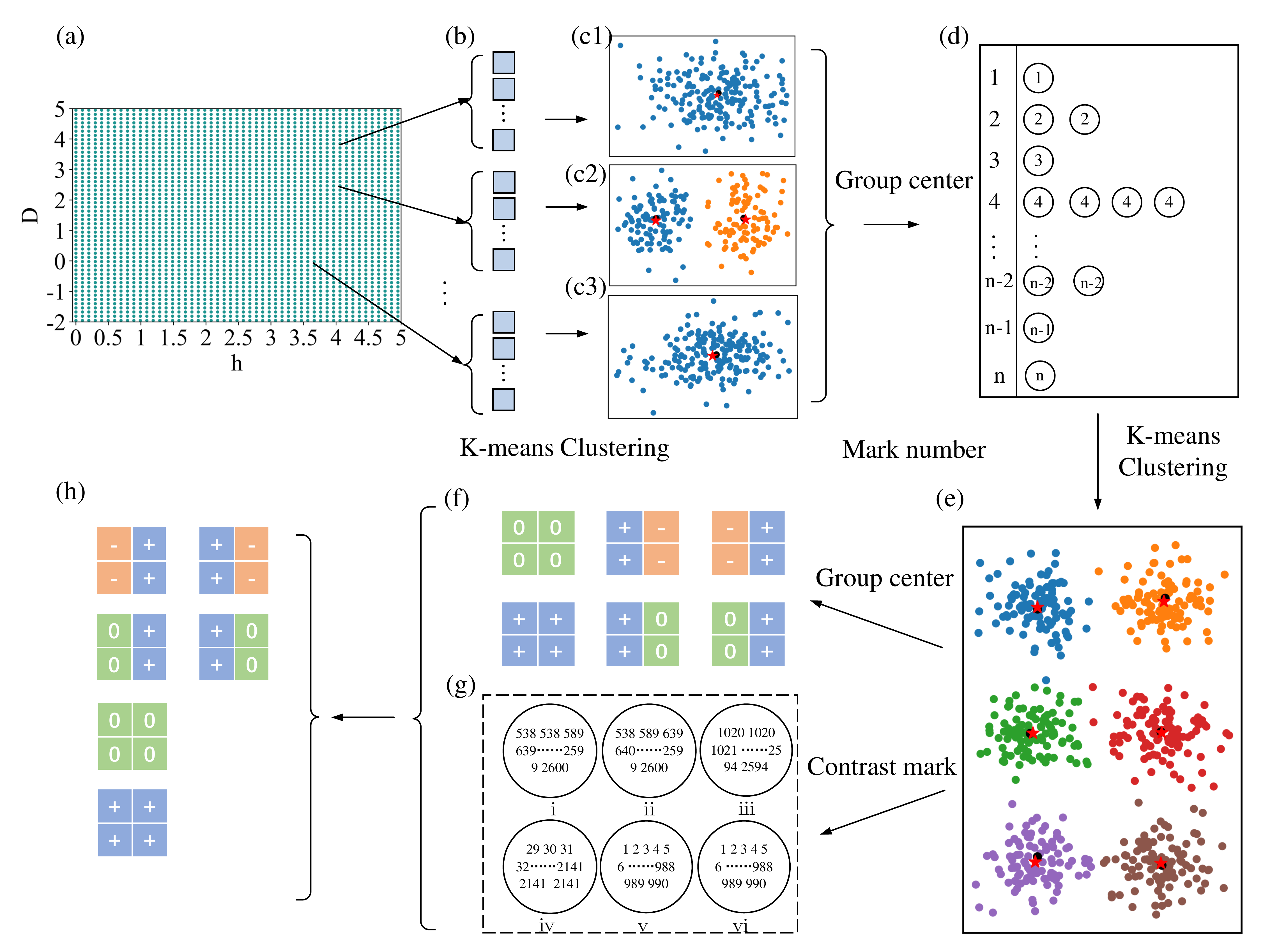}
\caption{The main idea of the two-times clustering method. (a) In the plane $(D, h)$, the parameter intervals are gridded in preparation for sampling. (b) For each parameter point, 100 configurations are sampled. (c) The first-time clustering determines the cluster centers for 100 samples at each parameter point. (d) The configurations with the same cluster center are given the same labels. (e) The second-time clustering determines the cluster centers of all samples in the whole parameter space. (f) and (g) The clusters are compared in pairs and similarity is judged based on the indexes of the clusters. (h) Groups with high similarity are considered to be in the same phase and given the same labels.}
\label{fig:Secondary Clustering Flowchart}
\end{figure*}

\section{Introduction}
Since the pioneering work of J. Carrasquilla and R. G. Melko on applying machine learning (ML) methods to study spin systems~\cite{Juan}, predicting the phase diagrams of interacting spin systems using ML has become an area of research interest~\cite{review}. Unsupervised ML methods are particularly useful in predicting phase diagrams as they do not require prior knowledge of the data labels. Common unsupervised ML methods include principal component analysis \cite{pca1, pca2, pca3}, t-distributed stochastic neighbor embedding~\cite{PhysRevE.97.013306, PhysRevB.103.075106}, diffusion maps~\cite{df1, df2, df3, PhysRevResearch.3.013074}, the confusion method~\cite{confusion}, the active contour model (snake model)~\cite{liuprl,sun2023snake}, Calinski-Harabaz  index\cite{chindex, prr}, and 
many others. Various indicators always signal at the phase transition point.

\setlength{\parskip}{4pt}
Sometimes, unsupervised methods incorporate certain steps of supervised learning. For example, the confusion method~\cite{confusion} determines labels to train the neural network (NN) based on guessed phase transition points. The best-guessed transition point is then determined based on the performance (accuracy) of the NN. In the NN-based snake model~\cite{liuprl, sun2023snake}, the learning network in the NN also needs to be trained by data and dynamical labels while cooperating with the other guessing network.

In contrast, Refs.~\cite{mlpotts} applied theoretical ground state configurations in the ordered phase as the training set to train the neural network without using real data, and the method was found to be effective for the Potts model. D.-R. Tan and F.-J. Jiang~\cite{PhysRevB.102.224434} even used datasets with only 0’s and 1’s to train the network and predict the critical points of the XY model and the $O(3)$ model. Indeed, the identification of ordered phases is crucial for predicting phase transitions accurately.

In this paper, we propose a two-times clustering method to identify ordered configurations and their corresponding labels from numerical simulations. In the first clustering step, we select representative configurations from each physical parameter point. In the second clustering step, we obtain perfect configurations and their corresponding labels. These selected perfected configurations and labels are then used to train the neural network, and the neural network is tested with real simulated data and mapped to a phase diagram.

The reason for performing the second clustering step is to prevent the neural network from misidentifying degenerate states as belonging to different phases. For example, this step of clustering avoids treating the ordered state configurations of the four lattice sites Ising model $\uparrow \uparrow \uparrow \uparrow $, $\downarrow \downarrow \downarrow  \downarrow $ as two different phases. The first step of classification involves selecting the perfect configurations from the data obtained through a real Monte Carlo (MC) simulation.

Our approach is essentially unsupervised despite utilizing neural networks for training and testing, as we do not know beforehand how many phases are included in the phase diagram. The $K$-means clustering algorithm~\cite{kmeans} is employed for clustering, which provides us with a group center or configuration. The number of perfect configurations is selected based on the criterion that the center of the cluster corresponds to the lowest energy configuration. In the second clustering step, the perfect configurations obtained from the first clustering step under all phase diagram parameter points are merged and clustered again.

We mainly choose the BC model~\cite{BC} to test our method. The BC model  on the square lattice is defined by the following Hamiltonian:
\begin{equation}
     	H=-J_{x}\sum_{<i,j>_{x}}S_{i}S_{j}-J_{y}\sum_{<i,j>_{y}}S_{i}S_{j} +D\sum_{i}S_{i}^2 -h\sum_{i}S_{i}  
      \label{eq:bch}
     \end{equation}
where $S_{i}= \pm 1,0, i=1,2 \cdots N$,  $N$ represents the total number of sites and $J_{x(y)}$ is  
the exchange interaction between sites along the two directions. $D$ is a single-spin anisotropy parameter and $h$ is an external magnetic field. 
The stripe-like pattern corresponds to the states  $\begin{smallmatrix} +-\\+- \end{smallmatrix}$, or $\begin{smallmatrix} +0\\+0 \end{smallmatrix}$.
The uniform pattern  corresponds to the states $\begin{smallmatrix} ++\\++ \end{smallmatrix}$, and $\begin{smallmatrix} 00\\00 \end{smallmatrix}$, respectively. The symbols ``+'', ``-'', and ``0'' correspond to the values taken by the spins. 
The phase diagram has been tested by the snake-net model~\cite{sun2023snake}.

By testing, we obtain two-dimensional planar phase diagrams of the BC model in an unsupervised manner, and even three-dimensional phase diagrams can be obtained. In order to understand the method, we have also illustrated the two-dimensional Ising model and Potts model~\cite{wu1982potts} as examples.

The outline of this paper is as follows. In Sec.~\ref{sec:method}, we present  the methods including the two-times clustering, total procedure, and network structure. The first time clustering obtains the perfect configuration at the cluster center with each parameter point and the second time clustering obtains configurations and labels at  the cluster centers with all parameter points.  And then the similarity test is performed to merge the similar configuration and assign them the same labels. 

In Sec.~\ref{sec:results},  the method is applied to the two-dimensional Ising, Potts, and BC models.
For the Ising and Potts models, instead of using two-time clustering, we directly use the 2*2   perfect configuration as the kernel matrix and thus calculate the similarity. For the BC model,  the perfect configuration and labels are used from the two-times clustering. With them, the similarity degree is obtained by the neural networks. Both the 2D and 3D phase diagrams are obtained.  
The conclusion and discussion are 
presented in Sec.~\ref{sec:con}. 

\section{Methods}
\label{sec:method}
\subsection{Two-times of clustering}
\label{sec:twostep}
\subsubsection{The first  clustering}
In Figs.~\ref{fig:Secondary Clustering Flowchart} (a)-(b), a two-dimensional parameter space is constructed with $D$-$h$ variables, where $D$ is chosen to belong to the range [-2,5] and $h$ to belong to [0,5]. The two dimensions are discretized by 51 points each, resulting in a total of $51\times 51$ parameter points. Monte Carlo simulations are then conducted for each parameter point ($D_i$, $h_i$), generating 100 spin configurations with a size of 16 $\times$ 16.

In Fig.~\ref{fig:Secondary Clustering Flowchart} (c), K-means clustering is performed on the 100 spin configurations generated at each parameter point. K-means clustering is an iterative unsupervised learning algorithm used for partitioning data into $n_c$ clusters. Given an input dataset ${\boldsymbol{D}}$ and the number of clusters, the algorithm outputs the classification centers and the corresponding cluster labels for each sample. The resulting clusters are typically presented in two-dimensional or high-dimensional plots, and the challenge is to determine the optimal number of clusters.

Here, some physical factors need to be considered. We try  the number of clusters in turn, say $n_c=1,2,3,4,\cdots$, and then compare the average energy $E_i, i\in [1,n_c]$ corresponding to the samples at the center of the clusters. The number of categories $n_c^{min}$ with the lowest energy is selected.
Usually the clustering center ${\mathbf X}_c$ given by Kmeans is not a real sample, marked by the red points in the scatter plot in Figs.~\ref{fig:Secondary Clustering Flowchart} (c).
We choose the true sample ${\mathbf X'}_c$ with the closest Euclidean distance to be the cluster center.

The purpose of selecting the cluster centers  is to pick out ordered ground states, such as  $\begin{smallmatrix} ++\\++ \end{smallmatrix}$, and $\begin{smallmatrix} 00\\00 \end{smallmatrix}$, respectively. 
In Figs.~\ref{fig:Secondary Clustering Flowchart} (c1) and (c2),
only one sample (represented by a red dot) and two samples (represented by red dots) are selected and passed on to the  data sets in Fig.~\ref{fig:Secondary Clustering Flowchart} (d).
In actual simulations or real experimental data, the data has thermodynamic  or quantum fluctuations. Or some possible meta-stable states where the system does not reach thermal equilibrium due to algorithmic limitations.




\subsubsection{The second  clustering}
In the second-time clustering, we
Perform   clustering on the samples in Fig.~\ref{fig:Secondary Clustering Flowchart} (d),  and   six clusters are obtained in Fig.~\ref{fig:Secondary Clustering Flowchart} (e) via K-means
lustering and  Calinski-Harabaz  index\cite{chindex}, which has been used for detecting topological phases\cite{prr}. The configuration at the center of each cluster is shown in Fig.~\ref{fig:Secondary Clustering Flowchart} (f), and each configuration can be identified by its corresponding identity number.

\begin{figure}[tbh]
\centering
\includegraphics[width=0.4\textwidth]{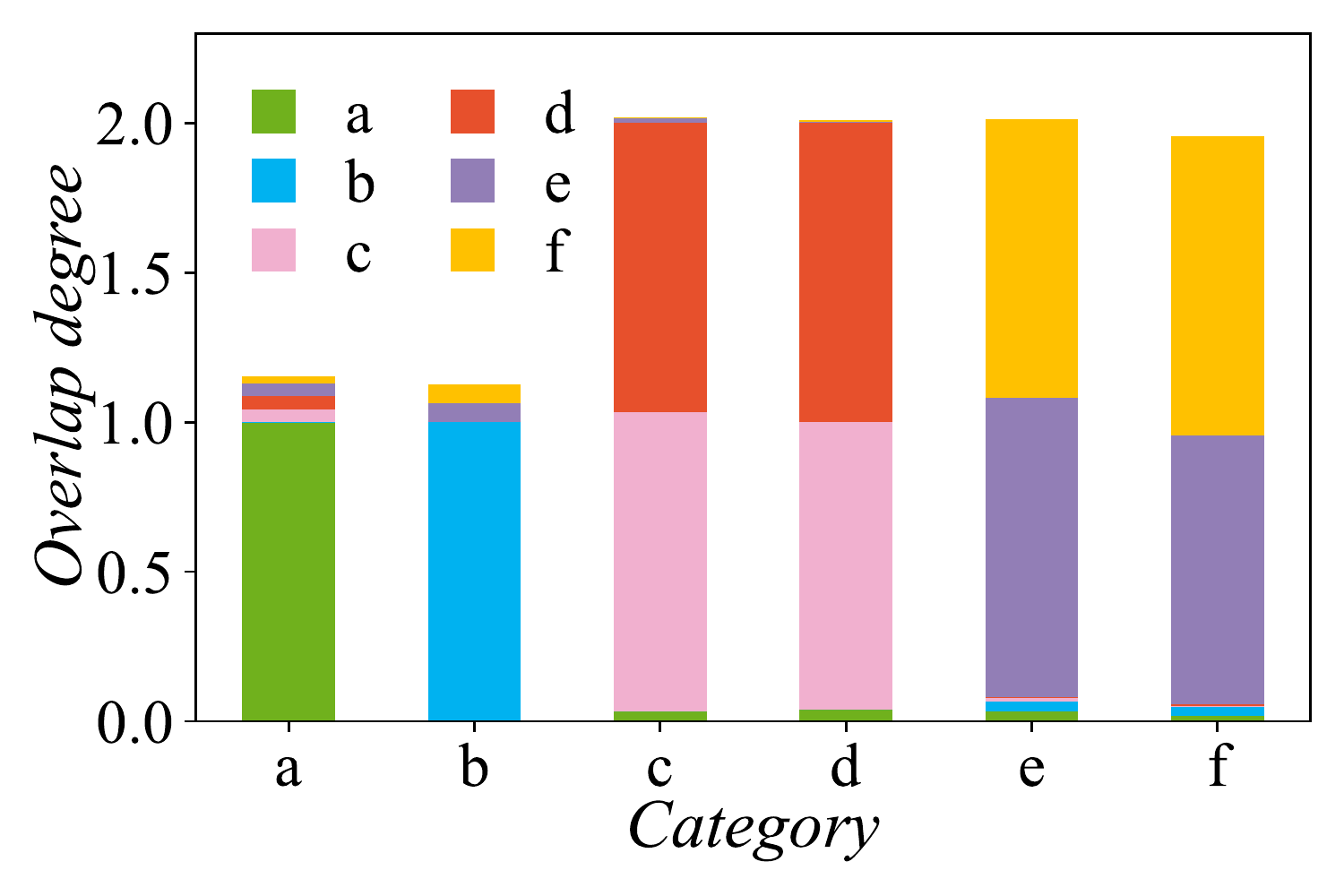}
\caption{Overlap degree versus classification}
\label{fig:overlap}
\end{figure}

\begin{figure*}[hbt]
\centering
\includegraphics[width=0.9\textwidth]{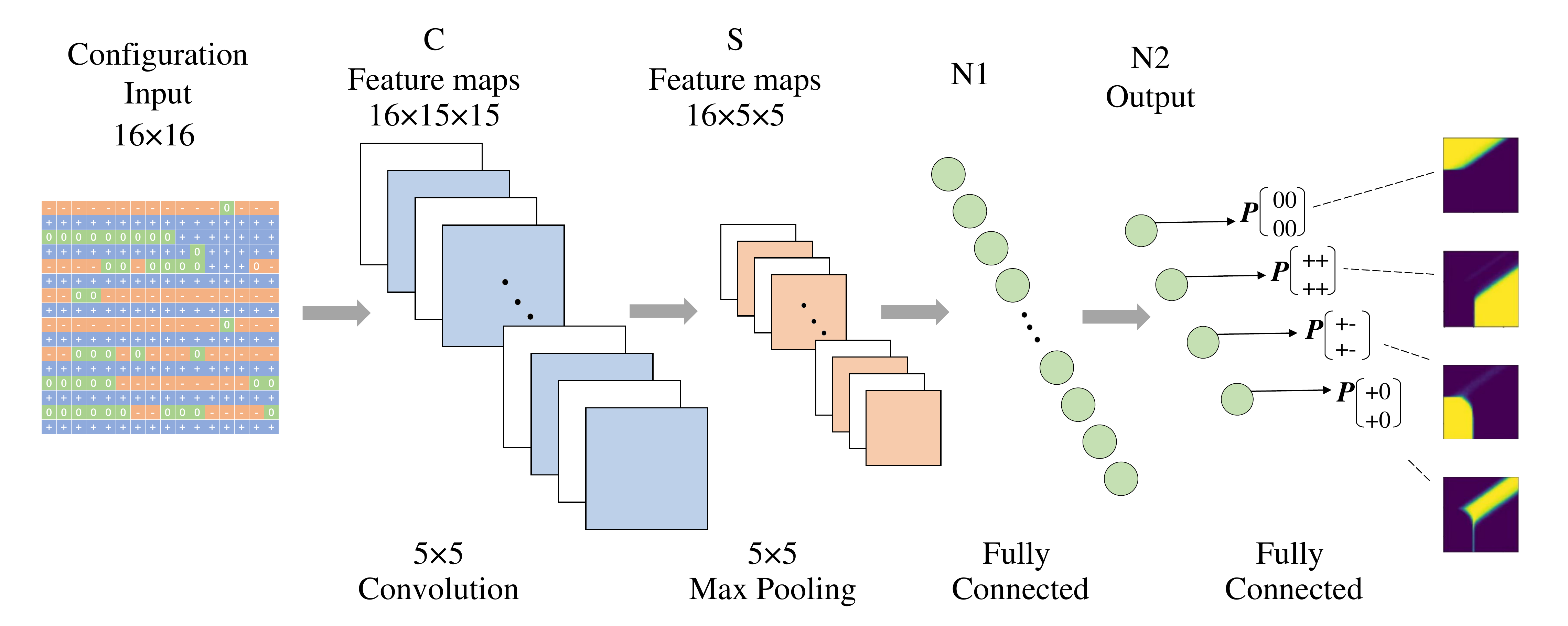}
\caption{The structure of the neural network. The input data size is 16*16. C: convolution layer, generating 16 feature maps of 15*15. S: pooling layer, generating 16 feature maps of 5*5. N1: The first fully connected layer with 16 neurons. N2: The output layer with 4 neurons. The output similarity degree $P$ is the average probability of the four categories.}
\label{fig:Neural_network_structure}
\end{figure*}

In Fig.~\ref{fig:Secondary Clustering Flowchart} (g), there are six clusters represented by circles, with each circle containing numerous ID numbers of the samples within the cluster. To measure the similarity between two clusters, we perform a similarity test separately for the ID numbers in each circle. The similarity measure used here is a commonly employed method in machine learning and statistics.

We first select a cluster $G_i$, and then search for all ID numbers in any of the groups $G_j=1,\cdots,6$. We count the number of overlapping ID numbers found and define the overlap degree $O_d(i,j)$ as the number of overlapping ID numbers between $G_i$ and $G_j$ divided by the total number of ID numbers in $G_i$. For example, if $G_i={1,2,3}$ and $G_j={2,3,4,5}$, then $O_d(i,j)=2/3$.

In Fig.~\ref{fig:overlap}, we observe the overlap degree between different categories calculated using the similarity measure as described earlier. For example, the overlap degree for a category with itself, $O_d(a,a)$, is marked by green color and has a value of 1. Additionally, we can see several small values for $O_d(a,j), j\ne a$, which represents the degree of overlap between the different categories.
Furthermore, categories $c$ and $d$ exhibit a high overlap degree of over 85$\%$, indicating that they are essentially the same category and should be grouped together.

\subsection{Total steps}
To obtain the phase diagram, we follow a three-step process:

(a) {\it Input data engineering.} Firstly, we use the Monte Carlo algorithm to generate spin configurations for the BC model on a two-dimensional 16 $\times$ 16 square lattice. These configurations are then assigned labels using the two-step clustering method illustrated in Fig.~\ref{fig:Secondary Clustering Flowchart}. The resulting labeled configurations $C_i$ are used in the subsequent steps.

(b) {\it Training and testing the neural network.} As shown in Fig.~\ref{fig:Neural_network_structure}, we use a convolutional neural network to classify the input configurations. The network is trained using the perfect configurations and labels from two-times clustering, and the entire data set is then input to the network for classification. The resulting classifications $P_i$ are recorded for each configuration.

(c) {\it Mapping out the phase diagram.} For each point $(x,y)$ in the parameter space, we calculate the gradients of the similarities $P_i$ with respect to $x$ and $y$ using the formulas:
\begin{align}
E_i(x,y) &= -|\nabla_xP_i|^2 -|\nabla_yP_i|^2, \\
E(x,y) & = \sum_{i=1}^{N_p} E_i(x,y),
\label{eq:energy}
\end{align}
where $N_p$ is the number of possible phases. The sum of the negative gradients provides information at each point in the parameter space. At the phase transition boundary, the change in similarity $P$ is the most significant, and consequently, the gradient is the largest. To extract the phase boundaries, we utilize a method inspired by the active contours model called the snake model~\cite{kass,liuprl,sun2023snake}.

The snake model employs energy minimization to identify contour or boundary lines in an image. Using this model, we define the energy of the image in Eq.~\ref{eq:energy}. This energy function enables us to trace and extract the phase boundaries in the parameter space.

\section{Results}
\label{sec:results}
\subsection{2D Ising and Potts models}

\begin{figure}[bth]
\centering
\includegraphics[width=0.45\textwidth]{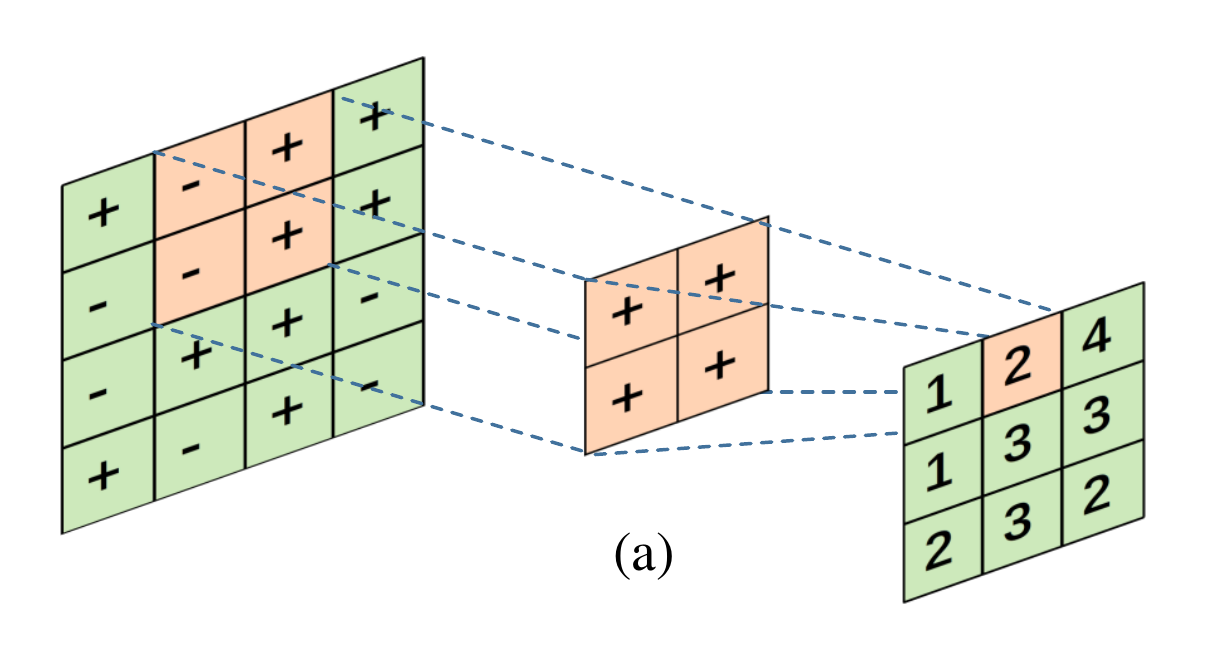}
\includegraphics[width=0.3\textwidth]{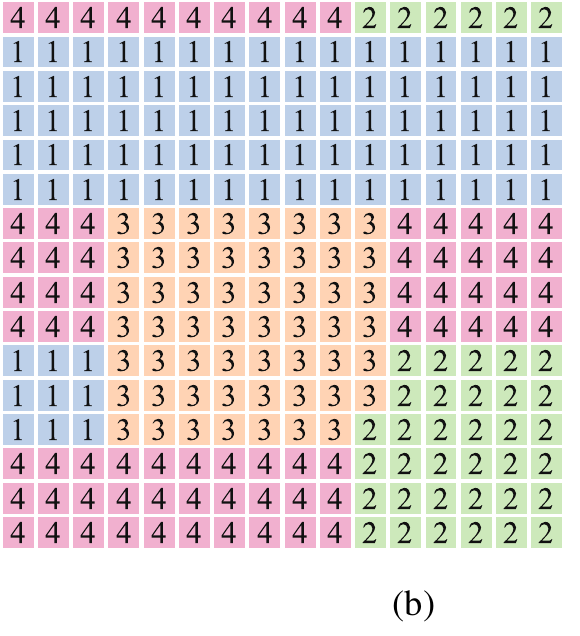}
\includegraphics[width=0.095\textwidth]{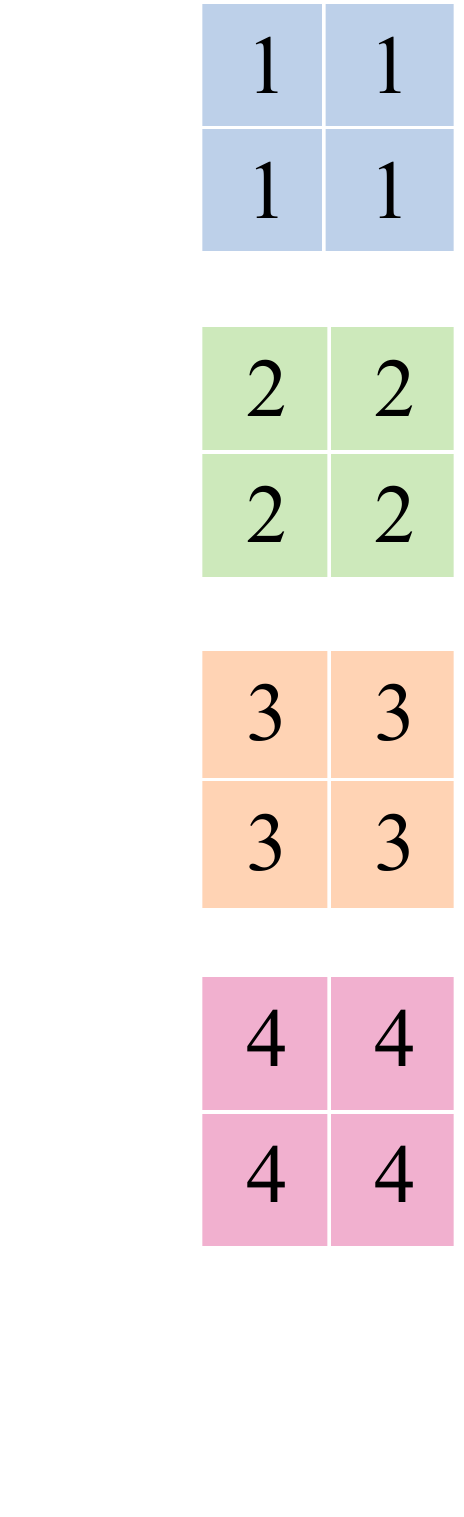}
\includegraphics[width=0.45\textwidth]{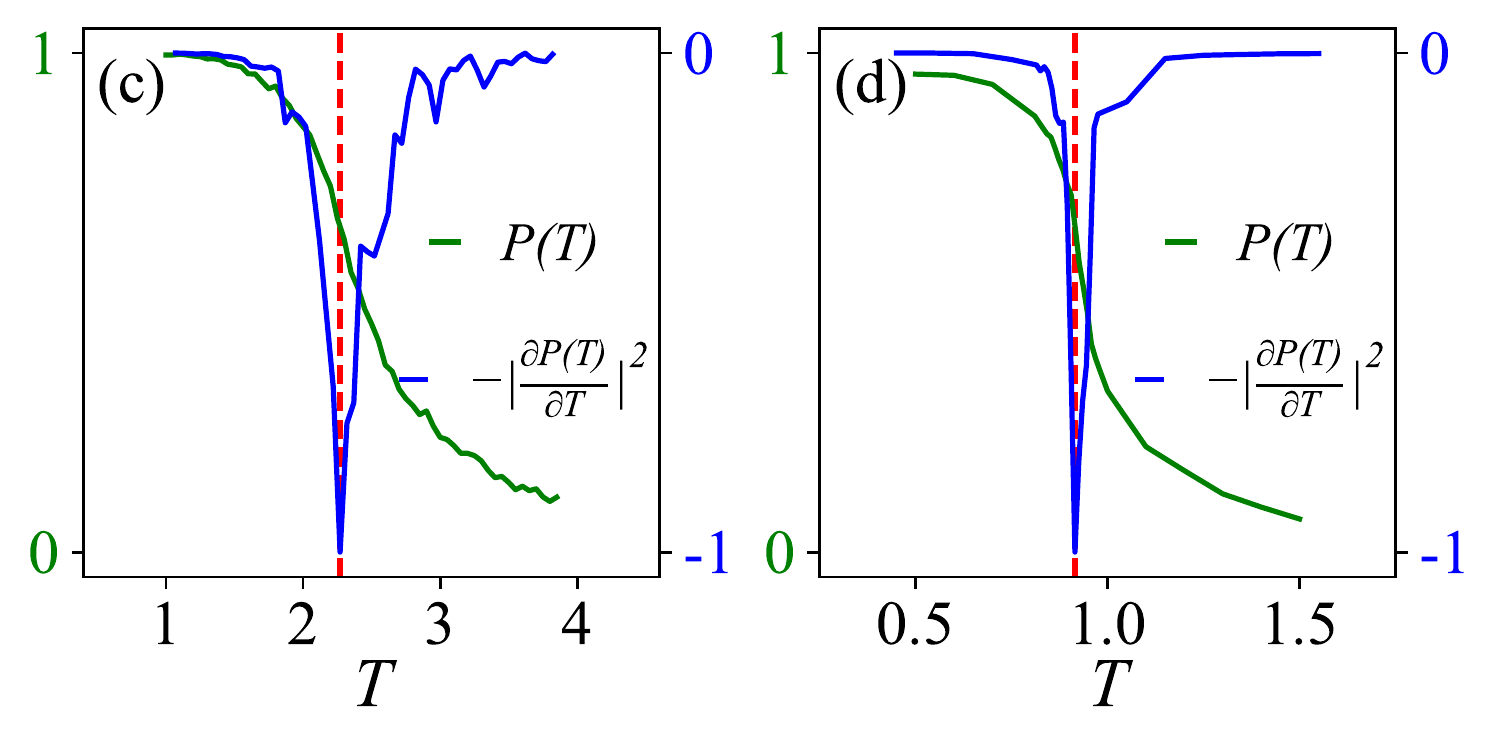}\par
\caption{(a) Schematic diagram of the XNOR operation. Comparing the input configuration and the kernel matrix element, the same is assigned a value of 1, while the difference gets a value of 0 and then local similarity $p_{i,j}$ is obtained. 
(b) A possible configuration of the 4-state Potts model and their kernels. 
The results $P(T)$ and $-|\frac{\partial P(T)}{\partial T}|^2$ from 
(c) the Ising model and (d) the Potts model. The red dashed line marks the location of the predicted phase transition point.}
\label{fig:fig_new_Ising_Potts_Phase_transition_point}
\end{figure}

The phase transition points of the 2D Ising model  and the Potts model~\cite{wu1982potts} are used as examples to  explain the meaning of similarity $P_i(x,y)$ and its image energy $E(x,y)$. Because only the temperature parameter is varied, the phase diagram is one-dimensional. At low temperatures, the ground state of the ferromagnetic system is very readily available. Instead of choosing the perfect configurations by clustering twice, they are given here directly.

As shown in Fig.~\ref{fig:fig_new_Ising_Potts_Phase_transition_point} (a), $P_i$ is obtained in a similar to  2D convolution.  For the Ising model, one first defines a  kernel matrix
as
\begin{equation}
    w^+=\begin{pmatrix}
1 & 1 \\
1 & 1
\end{pmatrix},
 \hskip 0.3cm
  w^{-}=\begin{pmatrix}
-1 & -1 \\
-1 & -1
\end{pmatrix}
\label{eq:two}
\end{equation}
and the kernel matrix element is all 1, representing a ground state a perfect configuration of the two-dimensional ferromagnetic Ising model. It is also feasible to use another perfect configuration that is all -1. Here, we use  two kernel matrices in Eq.~\ref{eq:two}.

The XNOR operation calculates the similarity degree of the kernel and the lattice by sliding the kernel through the entire lattice.  The similarity  for the Ising model  is defined as follows:
\begin{equation}
p_{i,j}^{+,-}=\sum_{m=1}^{kx}\sum_{n=1}^{ky} XNOR(w_{m,n}^{+,-},S_{i+m-1,j+n-1}).
\end{equation}
In this equation, $S_{i+m-1,j+n-1}$ represents the spin of the $(i+m-1)$-th row and $(j+n-1)$-th column of the input configuration. Additionally, $kx$ and $ky$ represent the width and height of the convolutional-like XNOR kernel, respectively. The function $XNOR(A,B)$ returns 1 if $A==B$ and 0 if $A\ne B$. The variable $p_{i,j}$ represents the local similarity degree of the $(i,j)$-th element of the lattice after the operation.


To calculate the total similarity, we first compare the local similarity $p_{i,j}^{+}$ and $p_{i,j}^{-}$ obtained from different kernels, take their maximum values and finally sum up to get the total similarity $P$, defined as 
\begin{align}
P_{Ising} &=\sum_{i,j}max(p_{i,j}^{+},p_{i,j}^{-}),\\
P_{q=4,Potts} &=\sum_{i,j}max(p_{i,j}^{1},p_{i,j}^{2},p_{i,j}^{3},p_{i,j}^{4}).
\label{eq:ppotts}
\end{align}

At temperature $T<T_c$, the 4-state Potts  model is likely to have the configuration shown in 
Fig.~\ref{fig:fig_new_Ising_Potts_Phase_transition_point} (b).
In the lattice, there are several blocks, and the spins are the same within each block. There are domain walls between the blocks,  caused by thermal excitation. This configuration is not the most perfectly ordered phase, e.g., all $1$, but it can still be ordered. For this purpose, we compare the similarity of $p_{i,j}^{i}$, $i=1,\cdots,4$, and take the maximum value as defined in Eq.~\ref{eq:ppotts} rather than take the average, which can not reflect the property of the phase very well.

 The results are shown in Fig.~\ref{fig:fig_new_Ising_Potts_Phase_transition_point} (c) and (d). For the Ising and Potts models, the similarity degree $P$ measures the similarity between the spin configurations.  $P(T)$  is marked by two green lines,  and $P(T)$  is high at low temperatures ordered phase  and decreases at high temperatures disordered phase.

To get the critical points, one has to measure the image energy defined by
 \begin{equation}
 E(T)=  -\sum_i |\frac{\partial P_i(T)}{\partial T}|^2.
\end{equation}
Actually, for the Ising model,
$T_{c}^{Ising}$  can be determined analytically  as:
\begin{equation}
T_c^{Ising}/J = 2/\ln(1+\sqrt{2}) \approx 2.269,
\end{equation}
where $J$ is the coupling constant between neighboring spins.  The critical behavior is reflected in the gradient of $P(T)$ with respect to temperature $T$, which exhibits a peak at $T_{c}^{Ising}$.
The critical temperature of the Potts model can be computed  by:
\begin{equation}
T_c^{Potts}/J = 1/{\ln(1+\sqrt{4})}=0.9102 .
\end{equation}
Similar to the Ising model, the critical temperature is marked by a peak in the gradient of $P(T)$.
The positions of the peak are located at 2.27 and 0.915, equal to the exact values within the intervals 0.01 respectively.
In summary, the similarity degree $P(T)$ and its derivatives enable us to identify the phase transition points.

\subsection{The Blume-Capel Model}

\begin{figure}
\centering
\includegraphics[width=0.4\textwidth]{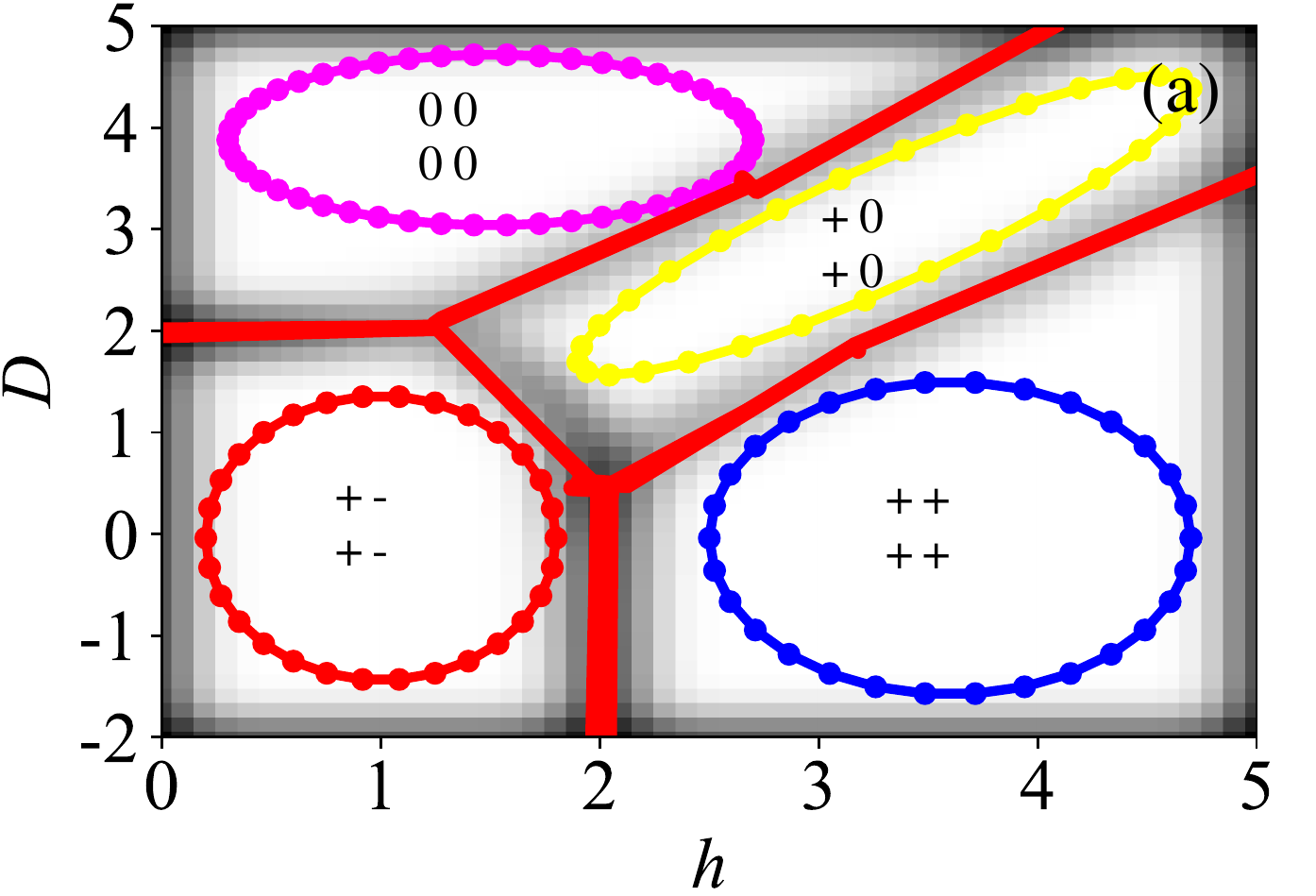}
\includegraphics[width=0.4\textwidth]{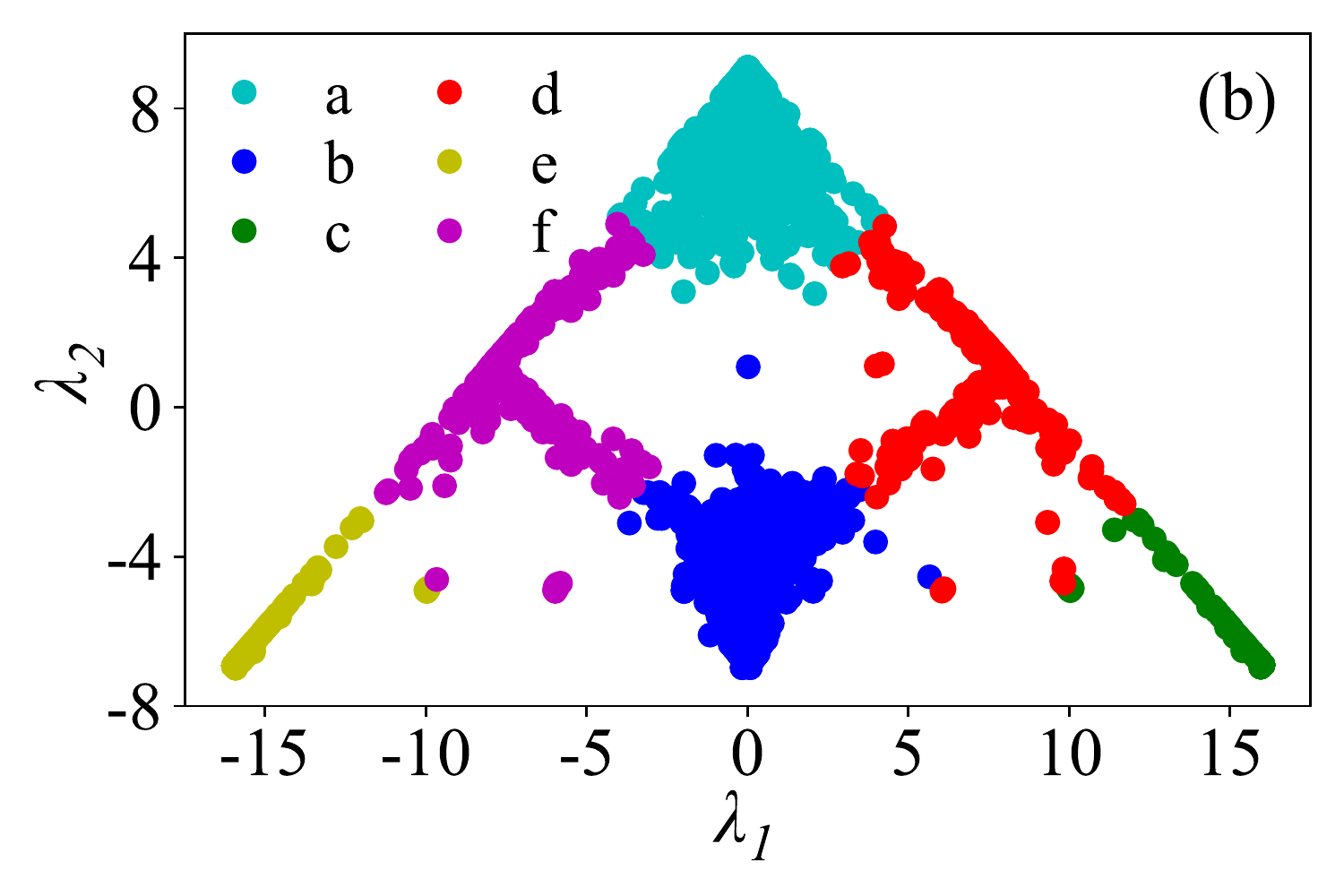}\par
\caption{ (a) Two-dimensional BC model, the image energy $E$ from the  gradient of similarity degree $P$ in the plane ($D-h)$.
(b) Real data corresponds to the results in Fig.~\ref{fig:Secondary Clustering Flowchart} (e). Six colors represent six clusters, pending a similarity test. The corresponding relation between each cluster and configuration are: a $\begin{smallmatrix} ++\\++ \end{smallmatrix}$,  b $\begin{smallmatrix} --\\-- \end{smallmatrix}$,  c $\begin{smallmatrix} -+\\-+ \end{smallmatrix}$,  d $\begin{smallmatrix} 0+\\0+ \end{smallmatrix}$,  e $\begin{smallmatrix} +-\\+- \end{smallmatrix}$ and 
 f $\begin{smallmatrix} +0\\+0 \end{smallmatrix}$. 
} 
\label{fig5:2dbc}
\end{figure}

\subsubsection{Two-dimensional phase diagram of the BC model}
The phase diagram of the BC model is known to be very rich~\cite{BC}. It has been shown to be a convenient model for testing machine learning algorithms~\cite{sun2023snake}. We use the proposed   method illustrated in Fig.~\ref{fig:Secondary Clustering Flowchart}. Firstly, we generate input data in the parameter space $D-h$ using the Monte Carlo method. Then, we carefully select the perfect configurations using our developed two-times clustering approach.

Finally, we train a convolution neural network (CNN) with the selected perfect configurations as input data and their corresponding labels and test the neural network to obtain predicted classifications.

According to Refs.~\cite{BC,sun2023snake}, in the low-temperature regime, the BC model exhibits four distinct phases. These phases have been classified based on their similarity degree, given by $P_1(\begin{smallmatrix} +-\\+- \end{smallmatrix})$, $P_2(\begin{smallmatrix} +0\\+0 \end{smallmatrix})$, $P_3(\begin{smallmatrix} ++\\++ \end{smallmatrix})$, and $P_4(\begin{smallmatrix} 00\\00 \end{smallmatrix})$, which are shown on the rightmost column of Fig.~\ref{fig:Neural_network_structure}.

In Fig.~\ref{fig5:2dbc} (a), the image energy (gray color) is obtained as
\begin{equation}
 E(D,h)=  -\sum_i |\frac{\partial P_i(D,h)}{\partial D}|^2 - |\frac{\partial P_i(D,h)}{\partial h}|^2.
\end{equation}
The four circular or elliptical contours denote the initial active contours, which move towards the boundary marked by the red solid lines under the influence of both image forces and internal forces~\cite{kass,sun2023snake}.

The scattering plot of the two-times clustering is also shown in Fig.~\ref{fig5:2dbc} (b).
The corresponding relations between each cluster and configuration are as follows, i.e., a $\begin{smallmatrix} ++\\++ \end{smallmatrix}$,  b $\begin{smallmatrix} --\\-- \end{smallmatrix}$,  c $\begin{smallmatrix} -+\\-+ \end{smallmatrix}$,  d $\begin{smallmatrix} 0+\\0+ \end{smallmatrix}$,  e $\begin{smallmatrix} +-\\+- \end{smallmatrix}$ and 
 f $\begin{smallmatrix} +0\\+0 \end{smallmatrix}$. Finally, by the similarity test as explained in Fig.~\ref{fig:overlap}, the cluster 
  c $\begin{smallmatrix} -+\\-+ \end{smallmatrix}$ and  e $\begin{smallmatrix} +-\\+- \end{smallmatrix}$ are overlapped and assigned with the same label. Similarly,  d $\begin{smallmatrix} 0+\\0+ \end{smallmatrix}$ and  f $\begin{smallmatrix} +0\\+0 \end{smallmatrix}$ are assigned with the  same labels. 
 
Looking closely at Fig.~\ref{fig5:2dbc} (b),  the scatter plot is symmetric about the  $\lambda_1=0$ axis due to the degenerate states.
These symmetrical patterns appear in diagrams in the Ref.~\cite{pca1,prr}, where the scattering of the generated ordered states ++++ and - - - -  of the Ising model are also distributed symmetrically. Here we use the similarity test to consider them as the same phase although they are separated in the scattering plot. 

In principle, the  phase behavior of the BC model can be obtained using the XNOR kernel method,
where the 2*2 small blocks are in the perfect configuration as kernels. In the neural network approach,  the perfect configurations and their labels are used  to train the kernel matrix.
The common goal is to obtain the similarity, but the XNOR method is simpler and easier to understand, and the neural network method is easier to extract the features of a large amount of data.
It is important to emphasize that in the neural network approach, the "labels"  are obtained through secondary clustering, and  they are unknown in advance. The method proposed in this paper is an unsupervised machine learning method.

\subsubsection{Three-dimensional phase diagram}
\begin{figure}
\centering
\includegraphics[width=0.3\textwidth]{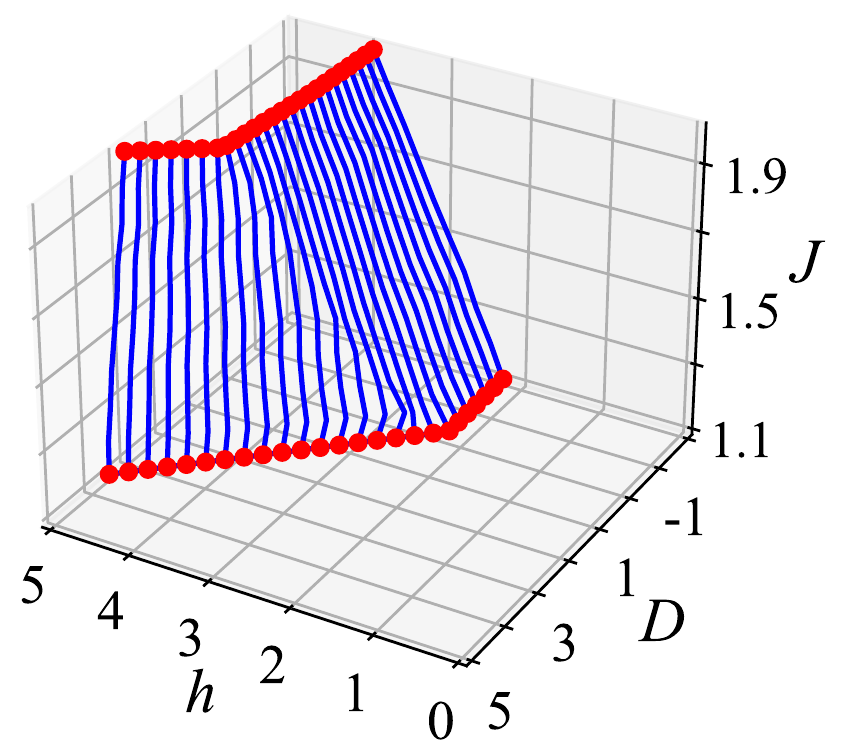}
\caption{The three-dimensional phase transition boundary of the phase  $\begin{smallmatrix} ++\\++ \end{smallmatrix}$.}
\label{fig:3d}
\end{figure}

In the Blume-Capel model, $J$ is a new variable introduced to form a three-dimensional phase diagram with $D$-$h$-$J$ as the three variables. However, to train the neural network, the 3D phase diagram is technically cut into 10 two-dimensional phase diagrams, $D$-$h$-$J_i$, where $J_i$ ranges between 1 and 2 in intervals of 0.1. As shown in Fig.~\ref{fig:3d}, the phase boundary of the $\begin{smallmatrix} ++
++ \end{smallmatrix}$ state shifts as $J$ varies.

Despite training with only one two-dimensional phase diagram, the neural network is able to generalize and accurately predict and classify all of the phase diagram data in the 10 two-dimensional plots. This is likely due to the fact that the patterns and structures of the phase diagram are similar across all values of $J$.

\begin{figure}
\centering
\includegraphics[width=0.4\textwidth]{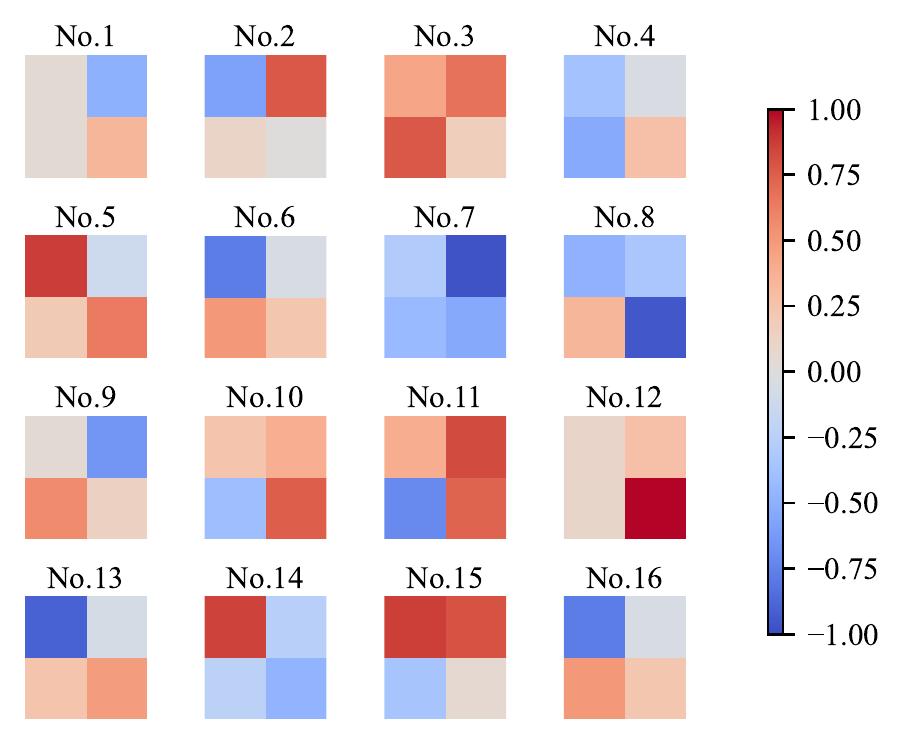}
\caption{Convolution kernels learned in the neural network.}
\label{fig:Convolution_kernel}
\end{figure}

The neural network trained using the perfect configuration obtained by two-times clustering can predict the phase diagrams corresponding to different $J$, and this is related to the fitting ability of the neural network. Fig.~\ref{fig:Convolution_kernel} shows the convolution kernels. The
various pattern is rich and exceeds the artificially defined ones. Unfortunately, clear interpretability is missing.  The data range of the kernel is almost symmetric, varying from -1 to 1, perhaps related to   the potential symmetry of the Hamiltonian or the spin variables of the BC model.

\section{Conclusion and Discussion}

In this paper, we propose a two-times clustering method to select perfect configurations from a range of degenerate samples. We use these configurations to train a neural network for the classification of two-dimensional phase diagram data and for identifying phase boundaries.
If we don't use our two-times clustering method, some dimensionality reduction algorithms may consider ++++ and - - - - ordered configurations as two separate phases and label them differently. 

We provide an example of the nearest-neighbor interaction Ising model or Potts model, where the temperature induces a phase transition from a ferromagnetic to a paramagnetic phase. Here, no two-times clustering is necessary to determine the phase transition point.
However, for more complex systems like the BC model containing rich phases, we need a two-times clustering method to select the perfect configurations and assign them the labels. 

While similar to using the Hamming distance to locate phase transition points in the literature~\cite{hamming}, this paper's primary focus is on two-times clustering to identify perfect configurations and assign the correct labels. 
Prior studies, such as Refs.~\cite{mlpotts}, used theoretical ground state configurations in ordered phases as a training set for neural networks.  For more complex systems, employing methods such as two-times clustering to identify perfect configurations is an alternative. 
Notably, we incorporate physics in the clustering center selection process, whereby the samples in the clustering centers correspond to the lowest energy.

In the future, we plan to test the robustness of the two-times clustering method by including examples such as frustrated lattices~\cite{Cewang}, quantum interacting bose-Hubbard systems~\cite{bh,jch}, topological systems~\cite{PhysRevB.102.224434}, and real material experiments.

{\it Acknowledgements:} We thank Lan Xinfei, Wang 
Xinzhu, Wang Yulin, Sun Xiaodong, and Ma Heyang for their assistance in this research. We gratefully acknowledge the support of the Hefei National Research Center for Physical Sciences at the Microscale (KF2021002) for this work.\\

\label{sec:con}
\bibliography{ref}
\end{document}